# Observation of second-order topological insulators in sonic crystals

Xiujuan Zhang[1#], Hai-Xiao Wang[2#], Zhi-Kang Lin[2], Yuan Tian[1], Biye Xie[1], Ming-Hui Lu[1,3,†], Yan-Feng Chen[1,3,], and Jian-Hua Jiang[2, †]

[1]National Laboratory of Solid State Microstructures and Department of Materials Science and Engineering, Nanjing University, Nanjing 210093, China

[2] School of Physical Science and Technology, and Collaborative Innovation Center of Suzhou Nano Science and Technology, Soochow University, 1 Shizi Street, Suzhou 215006, China

[3]Collaborative Innovation Center of Advanced Microstructures, Nanjing University, Nanjing 210093, China

[†]Correspondence and requests for materials should be addressed to joejhjiang@hotmail.com (JHJ) or

luminghui@nju.edu.cn (MHL)

*# These authors contributed equally to this work.*

**Topological insulators with unique gapless edge states have revolutionized the understanding of electronic properties in solid materials [1-4]. These gapless edge states are dictated by the topological invariants associated with the quantization of generalized Berry's phases of the bulk energy bands through the bulk-edge correspondence, a paradigm that can also be extended to acoustic and photonic systems [5-23]. Recently, high-order topological insulators (HOTIs) are proposed and observed, where the bulk topological invariants result in gapped edge states and in-gap corner or hinge states, going beyond the conventional bulk-edge correspondence [24-35]. However, the existing studies on HOTIs are restricted to tight-binding models and are incompatible with the conventional sonic/photonic crystals, where the energy bands are resulted from multiple Bragg scatterings. Here, we report theoretical prediction and experimental observation of acoustic second-order topological insulators (SOTIs) in two-dimensional (2D) sonic crystals (SCs) beyond the tight-binding picture. We observe gapped edge states and in-gap topological corner states which manifest the bulk-**

**edge correspondence in a hierarchy of dimensions. Moreover, topological transitions are discovered in both the bulk and edge by tuning a single geometric parameter, demonstrating topological control of acoustic waves in multiple dimensions. Our study opens up a new paradigm for topological designs of low-dimensional acoustic modes as well as topological transfer of acoustic energy among 2D, 1D and 0D states.**

Since the discovery of the quantum Hall effect, tremendous attention has been paid to topological phases of matters, as they go beyond the conventional classification of phases of matters based on spontaneous symmetry breaking, namely the well-known Ginzburg-Landau-Wilson paradigm [1-4]. Materials of the same symmetry can have distinct physical properties as characterized by the topological invariants of the energy bands below the band gap where the Fermi level resides. These topological invariants of an insulator result in robust gapless edge states and quantized physical responses (e.g., Hall conductance) [1-4]. Recently, by gapping out the edge states of topological insulators, HOTIs are created, where robust corner or hinge states emerge in the edge gap due to topological mechanisms [24-35].

It has been shown that the concept of topological insulators can be extended to acoustic waves in 2D SCs, leading to the discovery of 1D gapless edge states as waveguide channels robust against disorders [16-23]. The quest for HOTIs in SCs, however, is not straightforward, since the existing proposals for HOTIs are based on tight-binding models with designed hopping configurations, which are difficult to realize in conventional SCs. Moreover, acoustic band structures in conventional SCs are associated with multiple Bragg scatterings and are not able to be described by tight-binding theory since the Wannier functions are not exponentially localized [36].

In this Letter, we construct acoustic SOTIs through topological crystalline insulators [37] induced by gapping out a Dirac point at the Brillouin zone corner (i.e., the M point) of a 2D square-lattice SC. Topological edge states emerge at the boundary between two SCs with opposite Dirac masses. Since the band-degeneracy at the Dirac point is protected by the crystalline symmetries which are absent at the boundaries, the edge states have gapped spectrum, leading to 1D massive Dirac dispersions [14]. Remarkably, the Dirac masses of the 1D edge bands have opposite signs for the boundaries along the $x$ and $y$ directions, resulting in emergent Jackiw-Rebbi soliton states [38,39] localized at the corners shared by these boundaries (see Fig.1a).

To realize the above scenario, we design a 2D square-lattice SC with four meta-atoms in each unit-cell (Fig.1b). Each meta-atom is a photosensitive resin (a 3D printable material) block, which can be regarded as impenetrable for acoustic waves. The choice of four meta-atoms as a unit-cell is in fact a doubling of the primitive cell but a minimal unit-cell that is compatible with the supercells for the edge and corner states [40]. By rotating the angle $\theta$ of the four meta-atoms, Dirac points can be created and gapped out in different ways. The topological phase diagram of the bulk acoustic band is illustrated in Fig. 1c, which contains two different phases: the normal band gap (NBG) phase and the SOTI phase. The latter corresponds to parity inversion as illustrated in Fig. 1c, where the M and Γ points have opposite parity for the two bands below the acoustic gap (see Supplementary Information). The acoustic band structures for the NBG state with $\theta = -25°$, the phase transition point at $\theta = 0$, and the SOTI state with $\theta = 45°$, are given in Fig. 1d. The phase transition is signaled by an emergent four-fold degenerate Dirac point at the M point (see Supplementary Information for the details of the Dirac point).

The double degeneracy on the Brillouin zone boundary lines MX and MY can be understood as a consequence of the glide symmetries, $G_x := (x, y) \to (\frac{a}{2} + x, \frac{a}{2} - y)$ and $G_y = (x, y) \to (\frac{a}{2} - x, \frac{a}{2} + y)$. When combined with the time-reversal operation $T$, the anti-unitary symmetry operators $\Theta_j = G_j * T$ ($j = x, y$) enable double degeneracy at the Brillouin zone boundaries. For instance, $\Theta_x^2 \psi_{n,\vec{k}} = e^{ik_x a} \psi_{n,\vec{k}}$, where $\psi_{n,\vec{k}}$ is a Bloch wavefunction for the acoustic pressure field with $n$ and $\vec{k}$ the band index and wavevector, respectively. At the Brillouin zone boundary line MX, $\Theta_x^2 = -1$, holds for all acoustic bands, inducing fermion-like Kramers degeneracy. The four-fold degenerate Dirac point at the M point is a consequence of the crossing between two pairs of doubly degenerate bands: the even-parity bands (the degenerate $s$ and $d$ wave states) and the odd parity bands (the degenerate $p_x$ and $p_y$ wave states) (see Fig. 1c). The gap opening for such a Dirac point gives rise to physics similar to the quantum spin Hall insulators, as shown in Refs. [13,14,16]. In the Supplementary Information, we establish the connection between the acoustic topological crystalline insulator and the Bernevig-Hughes-Zhang model for quantum spin Hall effects [41] through a Hamiltonian theory. Such a connection is confirmed by the properties of the edge states: the edge states carry finite angular momentum (pseudo-spin) and exhibit pseudo-spin-wavevector locking as shown in Fig. 1e and Fig. 2 (see Supplementary Information for more details).

The two pseudo-spin polarized 1D edge bands of the topological crystalline insulator are supposed to cross each other at the $k_y = \frac{\pi}{a}$ ($k_x = \frac{\pi}{a}$) point to form gapless Dirac dispersions for the edge along the $y$ ($x$) direction. However, because the glide symmetries that protect the topological order are broken on those boundaries, the edge bands are generally gapped, leading to massive 1D Dirac dispersions. The Dirac mass at the edges, $m_y$ ($m_x$), can be characterized by the frequency difference between the odd and even modes at the $k_y = \frac{\pi}{a}$ ($k_x = \frac{\pi}{a}$) point.

Remarkably, these Dirac masses are also tunable through the rotation angle of the meta-atoms. For instance, for the boundaries between the SC with $\theta = -25°$ and the SC with $\theta = 45°$, the Dirac masses are of opposite signs for the edges along the $x$ and $y$ directions. This sign difference leads to the formation of the Jackiw-Rebbi soliton modes localized on the corners shared by the two edges. The emergence of these topological corner modes manifests bulk-edge correspondence in a hierarchy of dimensions: the bulk topology leads to the edge states, while the edge topology leads to the corner states, as shown in Fig. 1e. Such hierarchical bulk-edge correspondence is the smoking-gun signature of the HOTIs.

The physical properties of the helical edge states are studied for the SCs with $\theta = -45°$ and $\theta = 45°$ as an example. The choice of these two SCs is to minimize the band gap of the edge states. The edge dispersion is measured in our experiments by the Fourier-transformed acoustic pressure field scan method (see Methods). The measured edge dispersion agrees with the calculated dispersion within the precision of the pressure field scan method (see Fig. 2a). In addition, the spatial distribution of the acoustic pressure field of the edge states measured in the experiments also agrees with the simulation as shown in Figs. 2b and 2c. Calculation indicates that the phase profiles of the acoustic pressure fields for the edge states at finite wavevectors have intriguing properties: there are phase singularity points (i.e., phase vortices) around which the phase winds $\pm 2\pi$ (see Fig. 2d), indicating finite acoustic angular momentum. Moreover, these vortices have opposite phase winding directions for opposite wavevectors, indicating opposite angular momentum for opposite wavevectors. This feature is consistent with the pseudo-spin-wavevector locking for the edge states of quantum spin Hall insulators. Therefore, our observations confirm that the SOTI SC is a topological crystalline insulator mimicking the quantum spin Hall effect in acoustic systems. We corroborate the above observations with a Hamiltonian analysis for the edge states and simulations on selective launching of the edge states via sources with finite angular-momentum in the Supplementary materials. These analyses confirm a massive Dirac equation description of the edge states through the Hamiltonian of $H_j = v_j(k_j - \frac{\pi}{a})\sigma_z + m_j\sigma_y$ with $j = x, y$ for the edges along the $x$ and $y$ directions, where $\sigma_z = \pm 1$ denotes the pseudo-spin up and down states, respectively.

We then measure the corner states in a box-shaped geometry formed by the two SCs with $\theta = -25°$ and $\theta = 45°$ (see the inset of Fig. 3a). The frequencies of the eigen-modes are shown in Fig. 3a where four nearly degenerate corner modes emerging in the gap of the edge modes are clearly visible (see Supplementary Information for more details). The sample in experiments with the box-shaped supercell geometry is shown in Fig. 3b. Same experimental procedure as that used in measuring the edge states is performed (see Methods). The simulated and measured distributions of the acoustic pressure field for one of the four corner states are shown in Figs. 3c and 3d, respectively. Both of them indicate strongly localized 0D acoustic modes. The transmission measurements, which further demonstrate the responses of the corner states in the edge gap, are provided in the Supplementary Information, along with the observation of the corner states at the other corners. The agreements between the measurements and the simulations confirm the emergence of the 0D topological corner states within the edge gap. Detailed studies in the Supplementary Information reveal that these corner states are robust against various defects.

We now study the topological phase transitions on the edge, by fixing the angle of the meta-atoms of the NBG SC at the outside as $\theta_{out} = -25°$, while tuning the angle of the meta-atoms for the SOTI SC lying inside the box-shaped supercell, $\theta_{in}$. The band gaps for the edge states along the $x$ and $y$ directions are controlled by the rotation angle $\theta_{in}$. The edge phase diagram is shown in Fig. 4a where several transitions are discovered. The edge gap closing at $\theta_{in} = 0°$ and $\theta_{in} = 90°$ are caused by bulk band gap closing and bulk topological transitions. Remarkably, another transition at $\theta_{in} = 25°$ appears where the edge gap closes for both edges along the $x$ and $y$ directions when the bulk band gap remains finite. Both the edge Dirac masses, $m_y$ and $m_x$, switch signs at this transition point (see Fig. 4a). The bulk and edge bands at the transition point are presented in Fig. 4b. This transition indicates an important feature of the SOTIs that a sole topological transition at the edge bands can happen, while the topology of the bulk bands does not change.

Above discussed topological transitions give rise to various transitions among the bulk, edge and corner states, as indicated in Fig. 4c. In a box-shaped supercell calculation, the corner states studied in Fig. 3 merge into the bulk when $\theta_{in}$ approaches 90° due to the bulk topological transition. However, when $\theta_{in}$ approaches 25°, the corner states merge into edge states due to the edge gap closing and the topological transition in the edge bands. Further decreasing of the angle $\theta_{in}$ leads to a transition from edge states to bulk states caused by the bulk topological transition at $\theta_{in} = 0°$. Beyond the region of $0° < \theta_{in} < 90°$, the edge and corner states vanish as the inner SC becomes a NBG SC. These topological transitions can be used to transfer acoustic energy among bulk, edge and corner modes. Finally, we remark that for $\theta_{in}$ close to or smaller than 25°, the edge gap is too small to support distinguishable corner states, although the Jackiw-Rebbi mechanism remains valid.

Achieving topological transitions among the bulk, edge and corner modes in a single chip is of fundamental importance, which can enable topologically protected 1D and 0D modes for integrated acoustic devices as well as topologically induced acoustic energy transfer among different dimensions. Our study uncovers the underlying physics for such hierarchical topological transitions and opens a new path for future applications of topologically protected integrated acoustic devices and topological acoustic energy/information guiding for hierarchical dimensions. Furthermore, it raises new questions important for future fundamental sciences and applications: Is it possible to extend the scenario discovered in this work to photonic or other classical wave systems? Can one realize SOTIs or third-order topological insulators for 3D sonic or photonic crystals where topological hinge states emerge? What will happen if the corner or edge states are selectively coupled to resonators with gain or loss? Will HOTIs provide design facility for advanced integrated acoustic or photonic circuits? These important questions will drive the study on HOTIs to future science and applications.

## Methods

### Experiments

The present SC consists of blocks made of photosensitive resin (modulus 2765 MPa, density 1.3 g/cm$^3$). We utilize a stereo lithography apparatus to fabricate the samples: one with $w = 0.15a$ and $l = 0.5a$ for the edge state measurements, and, the other with $w = 0.1a$ and $l = 0.4a$ for the corner state measurements, where $a = 2$ cm is the lattice constant. The geometric tolerance is roughly 2 mm. The height of our samples is chosen to be 1 cm, roughly half the lattice constant. In the measurements, the samples are enclosed in the vertical direction by two flat acoustically rigid plates to form a waveguide, such that at the operating frequencies ~10 kHz, only 0th order mode can be excited. This ensures that the 2D approximation is applicable (see Supplementary Information for more details). The experimental data plotted in Fig. 2a are collected from the following procedure. We first scan the acoustic pressure field distribution under a specific frequency excitation. The experiments are conducted using an acoustic transducer to generate acoustic signals, which are guided into the samples from channels (with diameter of ~4mm) opened on the bottom of the waveguide. An acoustic detector (B&K-4939 1/4 inch microphone) is used to probe the excited pressure field from an open channel (with diameter slightly larger than the detector) on the top of the waveguide, and its position is controlled by an automatic stage. The data are collected and analyzed by a DAQ card (NI PCI-6251). The obtained real space data are then taken for the Fourier transform to obtain the dispersion of edge states in momentum space, where the Matlab built-in function *fft* is used. For the measurement of the edge and corner states field distributions (Figs. 2c and 3d), we use the abovementioned procedure to collect data, which are further post-processed to generate the color maps. The transmission measurements in the Supplementary Information are conducted using the same set-up, only we fix the automatic stage and sweep the input frequencies.

**Simulations**

Numerical simulations in this letter are all conducted using the 2D acoustic module of a commercial finite-element simulation software (COMSOL MULTIPHYSICS). The resin blocks are treated as acoustically rigid boundaries. In the eigen-evaluations, all four boundaries of the unit cell are set as Floquet periodic boundaries for the data in Figs. 1c and 1d. The boundaries of the supercells are set as Floquet periodic boundaries along the edge direction, with the perpendicular direction set as plane wave radiation boundaries, for the data in Figs. 1e, 2a, 2b, 2d, 4a and 4b. The boundaries of the corner samples are set as plane wave radiation boundaries, for the data in Figs. 3a, 3c and the acoustic pressure field distributions in Fig. 4c.

**Acknowledgements**

X.J.Z, Y.T, B.X and M.H.L are supported by the National Natural Science Foundation of China (Grant No. 11625418 and No. 51732006). H.X.W, Z.K.L and J.H.J are supported by the National Natural Science Foundation of China (Grant No. 11675116). X.J.Z thanks Hao Ge and Siyuan Yu for the useful discussions on the experimental measurements, and Zeguo Chen for the help on simulation set-up.

**Author contributions**

J.H.J and M.H.L conceived the idea and guided the research. X.J.Z, H.X.W and Z.K.L performed the numerical simulations. H.X.W, B.X and J.H.J did theoretical analyses. X.J.Z and Y.T performed experimental measurements. All the authors contributed to the discussion of the results and the manuscript preparation. J.H.J, X.J.Z, and M.H.L wrote the manuscript.

## Figures

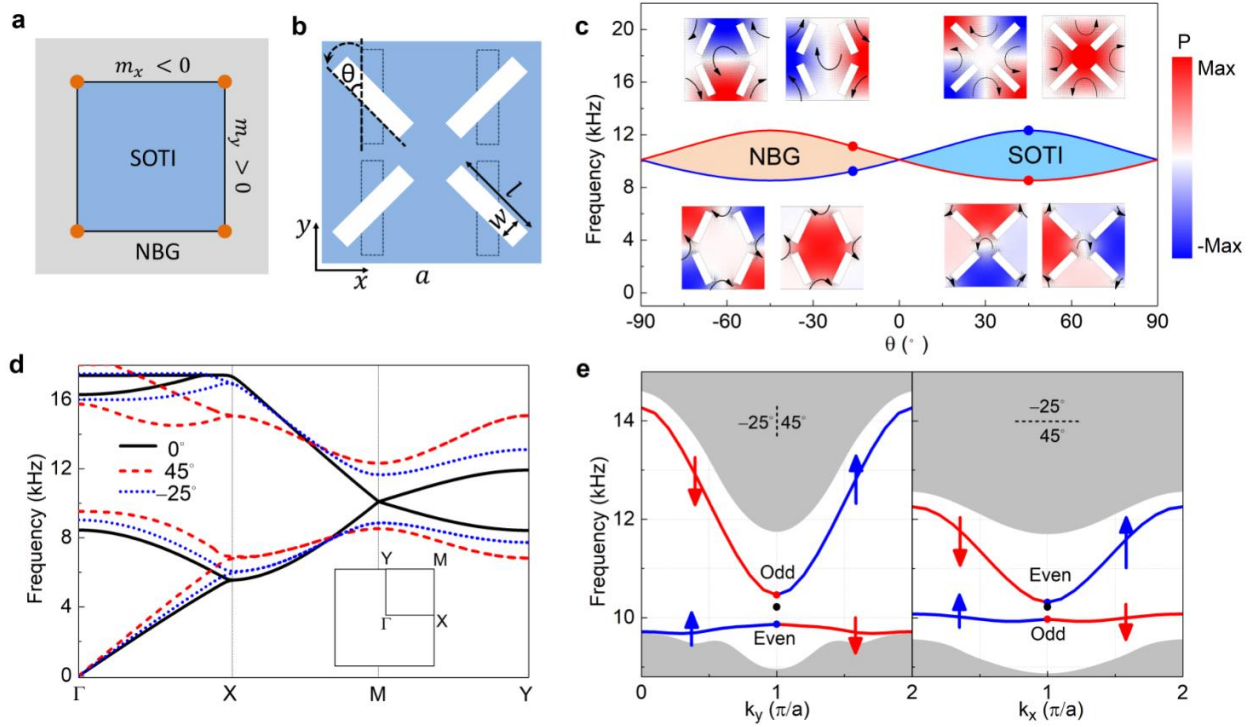

**Figure 1 | Bulk, edge and corner states in SOTIs. a,** Schematic for the corner states. The edge states on the *x* and *y* boundaries are 1D massive Dirac waves that carry Dirac masses with opposite signs, leading to topological corner states according to the Jackiw-Rebbi mechanism. **b,** Unit-cell of the proposed SC, composed of four meta-atoms that can be rotated. The rotation angle is denoted by $\theta$. **c,** Topological phase diagram of the bulk acoustic bands, represented by the frequencies of the odd (red curve) and even (blue curve) modes at the M point as $\theta$ is tuned. Insets: acoustic pressure fields of the bands above and below the gap at the M point for the angles labeled by the red and blue points. The black arrows indicate the Poynting vectors. **d,** Band structures of the SCs with $\theta = 0°$ (solid curves), $45°$ (dashed curves) and $-25°$ (dotted curves), with the inset depicting the Brillouin zone. **e,** The bulk (gray area), edge (solid curves) and corner (black dots) dispersions when the two SCs with $\theta = -25°/45°$ are placed together to form edges along the *y* (left) and *x*

(right) directions. The red and blue arrows label the pseudo-spin down and up, respectively. The parities of the edge states at $k_y = \frac{\pi}{a}$ and $k_x = \frac{\pi}{a}$ are also indicated in the figure.

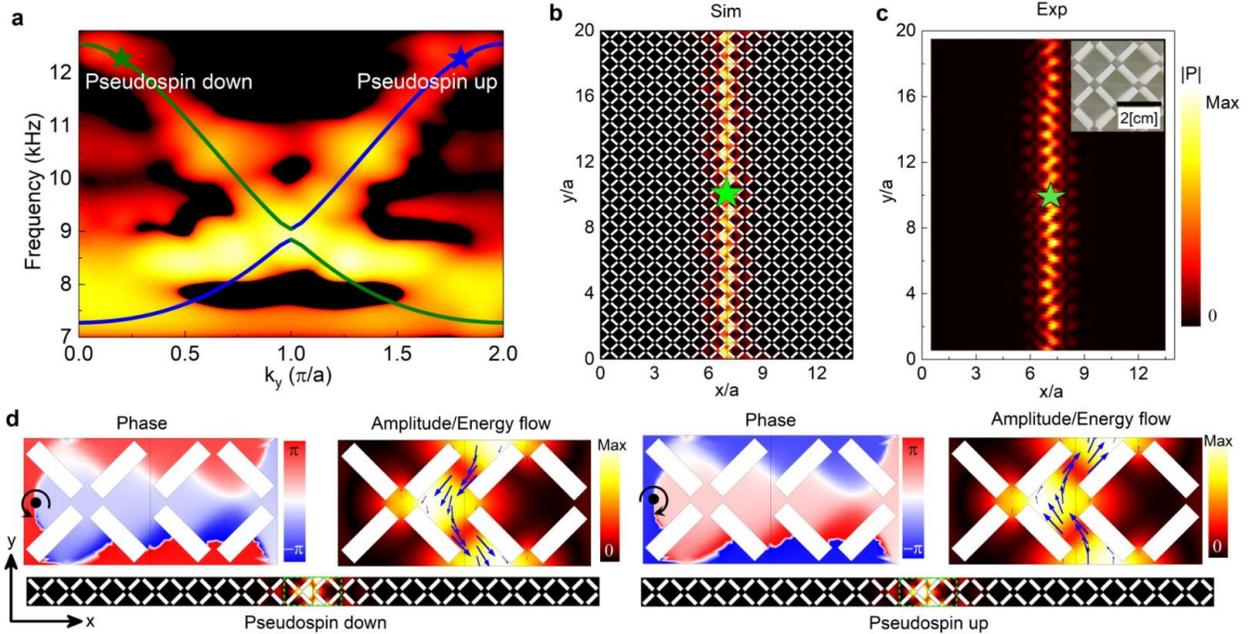

**Figure 2 | Characterizations of the edge states. a,** Simulated (blue and green curves) and measured (hot color) dispersions of the edge states. The measured data are obtained by taking Fourier transformation of the acoustic pressure field scanned along the edge at various frequencies. **b and c,** Simulated and measured acoustic pressure fields for the edge states launched by a point source (the green star), respectively. The inset depicts a part of the fabricated sample with the length scale. **d,** Acoustic pressure field distributions for the two edge states marked as (green and blue) stars in **a**, with the phase, amplitude and Poynting vectors (the blue arrows) shown for the two unit-cells close to the boundary. There are vortices in the phase distributions, indicating finite angular momentum (i.e., pseudo-spin). The vortex centers are labeled by the black dots, while the phase winding directions are indicated by the black arrows, for both the pseudo-spin up (right) and down (left) states.

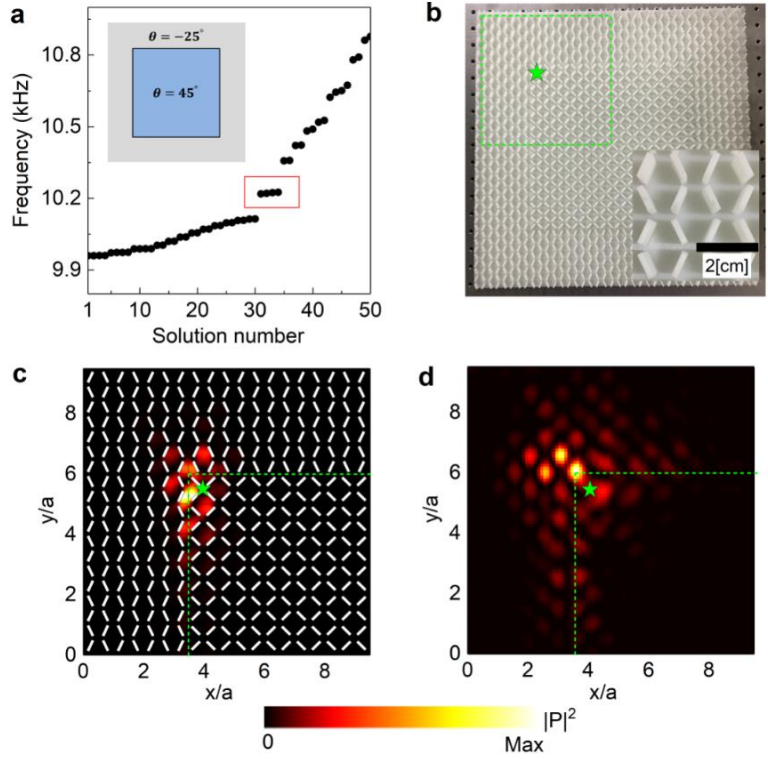

**Figure 3 | Measurements of the corner states. a,** Frequencies of the acoustic eigen-modes for a box-shaped supercell calculation (illustrated by the inset). The supercell consists of an area of $12 \times 12a^2$ SC with $\theta_{in} = 45°$, enclosed by a wall of SC with thickness $4a$ and $\theta_{out} = -25°$. There are four degenerate corner states in the edge gap, as highlighted by the red box. **b,** Image of the supercell sample, fabricated for the measurement of the corner states. The green star represents the point source used to excite the corner state. The inset illustrates the zoom-in structure of the outside SC and the length scale. **c and d,** The distributions of the acoustic pressure field of the corner state from the simulation and the experiment, respectively. For the sake of clarity, only the upper-left quarter (the green block in **b**) of the sample is displayed, with the *x* and *y* edges guided by the green lines.

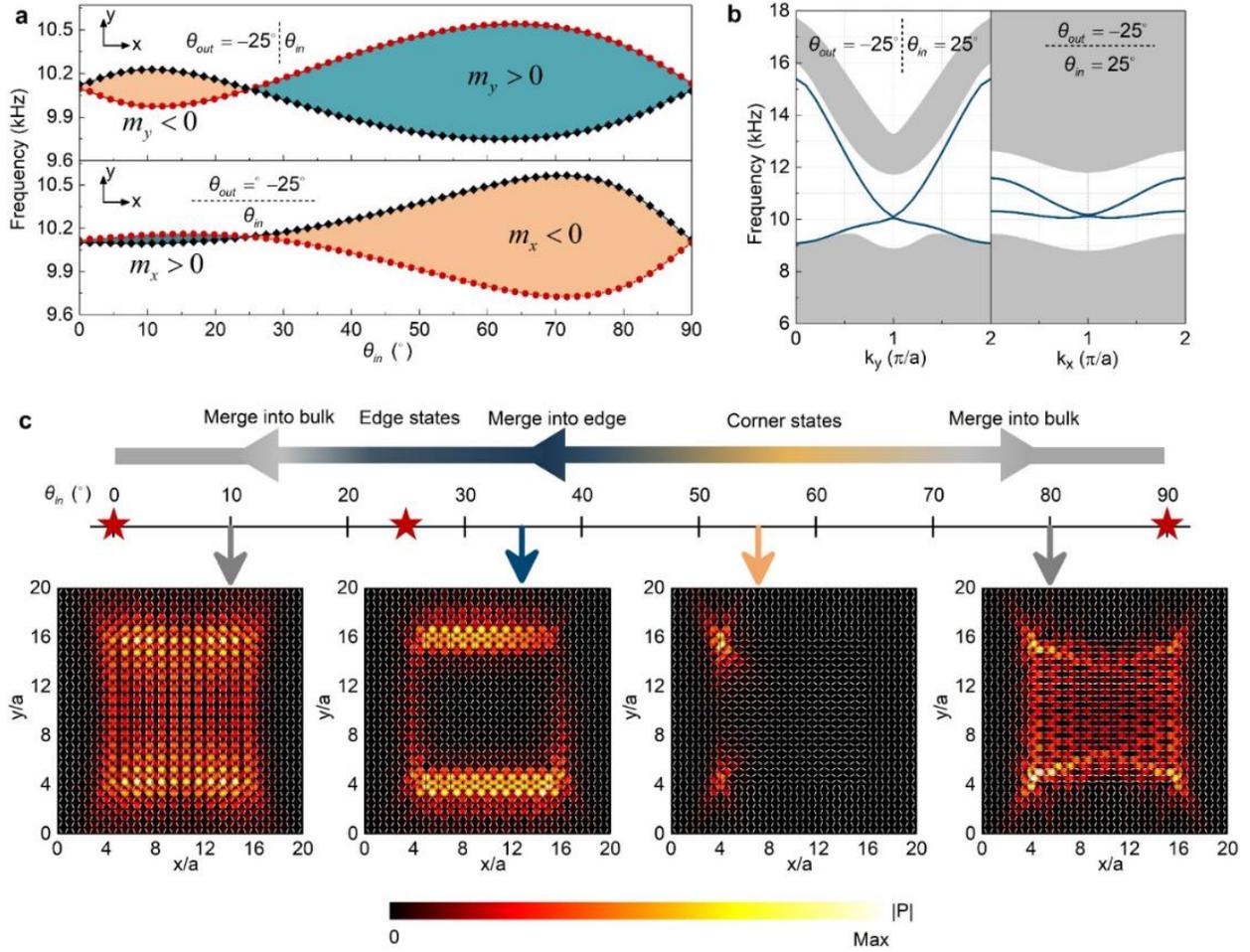

**Figure 4 | Topological transitions in the edge. a,** Topological phase diagram for the edge states, represented by the frequencies of the odd (red dots) and even (black rhombus) modes at $k_j = \frac{\pi}{a}$ with $(j = x, y)$. The supercells used to obtain these data are constructed by stacking together the NBG SC with $\theta_{out} = -25°$ and the SOTI SC with various $\theta_{in}$. The signs of the Dirac masses $m_x$ (on the *x* edge) and $m_y$ (on the *y* edge) mark various edge topological transitions. Two of them are at $\theta_{in} = 0°$ and $90°$, which are associated with the bulk topological transitions. Another one is at $\theta_{in} = 25°$, which is solely due to the edge property. **b,** The projected band structures for the supercell with $\theta_{out}/\theta_{in} = -25°/25°$. The edge gap closing is observed for both edges along the

$y$ (left) and $x$ (right) directions, indicating an edge topological transition. **c,** Demonstration of various transitions between the corner, edge and bulk states, induced by the edge topological transitions as $\theta_{in}$ is tuned. Such a process is sketched by the arrows and the gradient colors, with the red stars marking the three transition points in **a**. To further visualize these transitions, we present in each topological phase the acoustic pressure field distribution of a representative eigenstate, which displays a bulk state, an edge state, a corner state and a bulk state again, successively from the left to right.